\documentclass{article} 
\usepackage{iclr2018_conference,times}
\usepackage{hyperref}
\usepackage{url}
\usepackage{graphicx}
\usepackage{enumitem}
\usepackage{multirow}
\usepackage{algorithm}
\usepackage{caption}
\usepackage{subfig}
\usepackage[noend]{algpseudocode}
\usepackage{amsmath,amssymb}
\title{Melody Generation for Pop Music via \\Word Representation of Musical Properties}

\author{Andrew Shin, L\'eopold Crestel$^\ast$, Hiroharu Kato, Kuniaki Saito, Katsunori Ohnishi, \\ 
	\bf Masataka Yamaguchi, Masahiro Nakawaki$^\dagger$, Yoshitaka Ushiku, Tatsuya Harada\\
THE UNIVERSITY OF TOKYO\\
$^\ast$IRCAM \\
$^\dagger$INNOVATION CREATIVE, INC. \\
\texttt{\{andrew,kato,k-saito,ohnishi,yamaguchi,}\\
\texttt{ushiku,harada\}@mi.t.u-tokyo.ac.jp} \\
$^\ast$\texttt{leopold.crestel@ircam.fr} \\
$^\dagger$\texttt{nakawaki.ici@gmail.com} \\
}

\iclrfinalcopy 

\begin{document}

\maketitle

\begin{abstract}
Automatic melody generation for pop music has been a long-time aspiration for both AI researchers and musicians. However, learning to generate euphonious melody has turned out to be highly challenging due to a number of factors. Representation of multivariate property of notes has been one of the primary challenges. It is also difficult to remain in the permissible spectrum of musical variety, outside of which would be perceived as a plain random play without auditory pleasantness. Observing the conventional structure of pop music poses further challenges. In this paper, we propose to represent each note and its properties as a unique `\textit{word},' thus lessening the prospect of misalignments between the properties, as well as reducing the complexity of learning. We also enforce regularization policies on the range of notes, thus encouraging the generated melody to stay close to what humans would find easy to follow. Furthermore, we generate melody conditioned on song part information, thus replicating the overall structure of a full song. Experimental results demonstrate that our model can generate auditorily pleasant songs that are more indistinguishable from human-written ones than previous models.
\footnote{https://github.com/mil-tokyo/NeuralMelody}
\end{abstract}

\setlength{\abovedisplayskip}{3pt}
\setlength{\belowdisplayskip}{3pt}

\section{Introduction}
Recent explosion of deep learning techniques has opened up new potentials for various fields of multimedia. Vision and language have been its primary beneficiary, particularly with rising interest in generation task. Considerable amount of recent works on vision and language have hinged beyond mere generation onto artistic aspects, often producing works that are indistinguishable from human works (\cite{GAN}; \cite{Radford}; \cite{ghost}). On the other hand, it is only recently that deep learning techniques began to be applied to music, and the quality of the results are yet far behind those in other domains, as there are few works that demonstrate both euphonious sound and structural integrity that characterize the human-made musical contents. This unfortunate status holds true for both music in its physical audio format and its abstraction as notes or MIDI (Musical Instrument Digital Interface).

Such lagging of deep learning-enabled music generation, particularly in music as abstraction, can be attributed to a number of factors. First, a note in a musical work contains various properties, such as its position, pitch, length, and intensity. The overall tendency of each property and the correlation among them can significantly vary depending on the type of music, which makes it difficult to model. Second, the boundary between musical creativity and plain clumsiness is highly indefinite and difficult to quantify, yet exists. As much as musical creativity cannot be limited, there is yet a certain aspect about it that makes it sound like (or not sound like) human-written music. Finally, music is not merely a series of notes, but entails an overall structure of its own. Classical music pieces are well-known for their high structural complexity, and much of pop music follows the general convention of verse - pre-chorus - chorus structure. This structure inevitably necessitates different modeling of musical components; for example, notes in the chorus part generally tend to be more high-pitched. It goes without saying that these structure-oriented variations further complicate the modeling of music generation.

In this paper, we propose a new model for music generation, specifically symbolic generation of melodies for pop music in MIDI format. The term ``\textit{pop music}" can have different meanings depending on the context, but we use the term in this paper to refer to its musical characteristics as conventionally accepted. Specifically, it refers to the songs of relatively short lengths, mostly around 3 minutes, with simple and memorable melodies that have relatively low structural complexity, especially in comparison to classical music. Music in MIDI format (or, equivalently, in notes) can be considered a discrete abstraction of musical sound, analogous to the relationship between text and speech. Just as understanding text is not only essential in its own merit, but provides critical clues to speech and language in general, understanding music at its abstraction can provide an ample amount of insights as to music and sound as a physical format, while being fun and significant \textit{per se}.

We address each of the challenges described above in our proposed model. First, we propose to treat a note and its varying properties as a unique `word,' as opposed to many previous approaches that took each property into consideration separately, by implementing different layers for generation. In our model, it suffices to train only one model for generation, as each `word' is an incarnation of all of its properties, thus forming a melody as a `sentence' consisting of those notes and the properties. This approach was inspired by recent successes in image captioning task (\cite{Karpathy}; \cite{ShowAndTell}; \cite{ShowAttendTell}), in which a descriptive sentence is generated with one word at a time in a recurrent manner, while being conditioned on the image features. Likewise, we generate the melody with one note at a time in a recurrent manner. The difference is that, instead of image features obtained via convolutional neural networks (CNN), we condition the generation process on simple two-hot vectors that contain information on chords sequences and the part within the song. Chord sequences and part annotations are automatically generated using multinomial hidden markov model (HMM) whose state transition probabilities are computed from our own dataset. Combining Bayesian graphical models with deep neural netweorks (DNN) has become a recent research interest (\cite{gal}), but our model differs in that HMM is purely used for feature input generation that is processed by neural networks.

Second, we enforce regularization policy on the range of notes. Training with a large amount of data can lead to learning of excessively wide range of pitches, which may lead to generation of melodies that are not easy to sing along. We alleviate these problem by assigning a loss function for the range of notes. Finally, we train our system with part annotation, so that more appropriate melody for the corresponding part can be generated, even when the given chord sequences are identical with other parts of the song. Apart from the main model proposed, we also perform additional experiments with generative adversarial networks (\cite{GAN}) and with multi-track songs.

Our main contributions can be summarized as following:
\begin{itemize}[noitemsep,nolistsep]
	\item proposal of a model to generate euphonious melody for pop music by treating each note and its properties as single unique ``word", which alleviates the complexity of learning
	\item implementation of supplementary models, such as chord sequence generation and regularization, that refine the melody generation
	\item construction of dataset with chord and part annotation that enables efficient learning and is publicly available.
\end{itemize}

\section{Related Works}
Most of the works on automatic music composition in the early days employed rule or template-based approach (\cite{rule1}; \cite{rule2}). While such approaches made important contributions and still continue to inspire the contemporary models, we mainly discuss recent works that employed neural networks as a model of learning to compose music, to make close examination and comparison to our model.

DeepBach (\cite{DeepBach}) aims to generate Bach-like chorale music pieces by employing pseudo-Gibbs sampling. They discretize time into sixteenth notes, generating notes or intervals at each time step. This marks a contrast to our model that does not have to be aware of each discrete time step, since positional information is already involved in the note representation. They also assume that only one note can be sung per instrument at a given time, dividing chords into different layers of generation. On the other hand, our model can handle multiple notes at the same position, since sequential generation of notes does not imply sequential \textit{positioning} of notes. As we will see in Section 4.3, our model can generate simultaneous notes for a single instrument. \cite{Magenta2} also take a similar approach of applying Gibbs sampling to generate Bach-like chorale music, but mostly share the same drawbacks that make a contrast to our model.

\cite{Magenta1} proposed RL Tuner to supplement recurrent neural networks with reinforcement learning by imposing cross-entropy reward function along with off-policy methods from KL control. \textit{Note RNN} trained on MIDI files is implemented to assign rewards based on the log probability of a note given a melody. They defined a number of music-theory based rules to set up the reward function. Our model, on the other hand, does not require any pre-set rules, and the outcome can be easily controlled with simple regularizations.

\cite{SongFromPi} proposed a hierarchical recurrent neural network model to produce multi-track songs, where the bottom layers generate the melody and the higher levels generate the drums and chords. They built separate layers for pitch and duration that generate an output at each time step, whereas our model needs only one layer for pitch and duration and does not have to be aware of time step. They also conditioned their model on scale types, whereas we condition our model on chord sequence and part information.


While generating music as physical audio format is out of scope of this paper, we briefly discuss one of the recent works that demonstrated promising results. Originally designed for text-to-speech conversion, WaveNet (\cite{WaveNet}) models waveform as a series of audio sample $x_t$ conditioned on all previous timesteps, whose dependence is regulated by causal convolutional layers that prevent the violations in ordering. When applied to music, it was able to reconstruct the overall characteristics of corresponding music datasets. While only for a few seconds with frequent inconsistency, it was able to generate samples that often sound harmonic and pleasant.

\begin{figure}[t]
	\centering
	\includegraphics[clip, trim=0cm 0cm 0cm 0cm, width=0.85\linewidth]{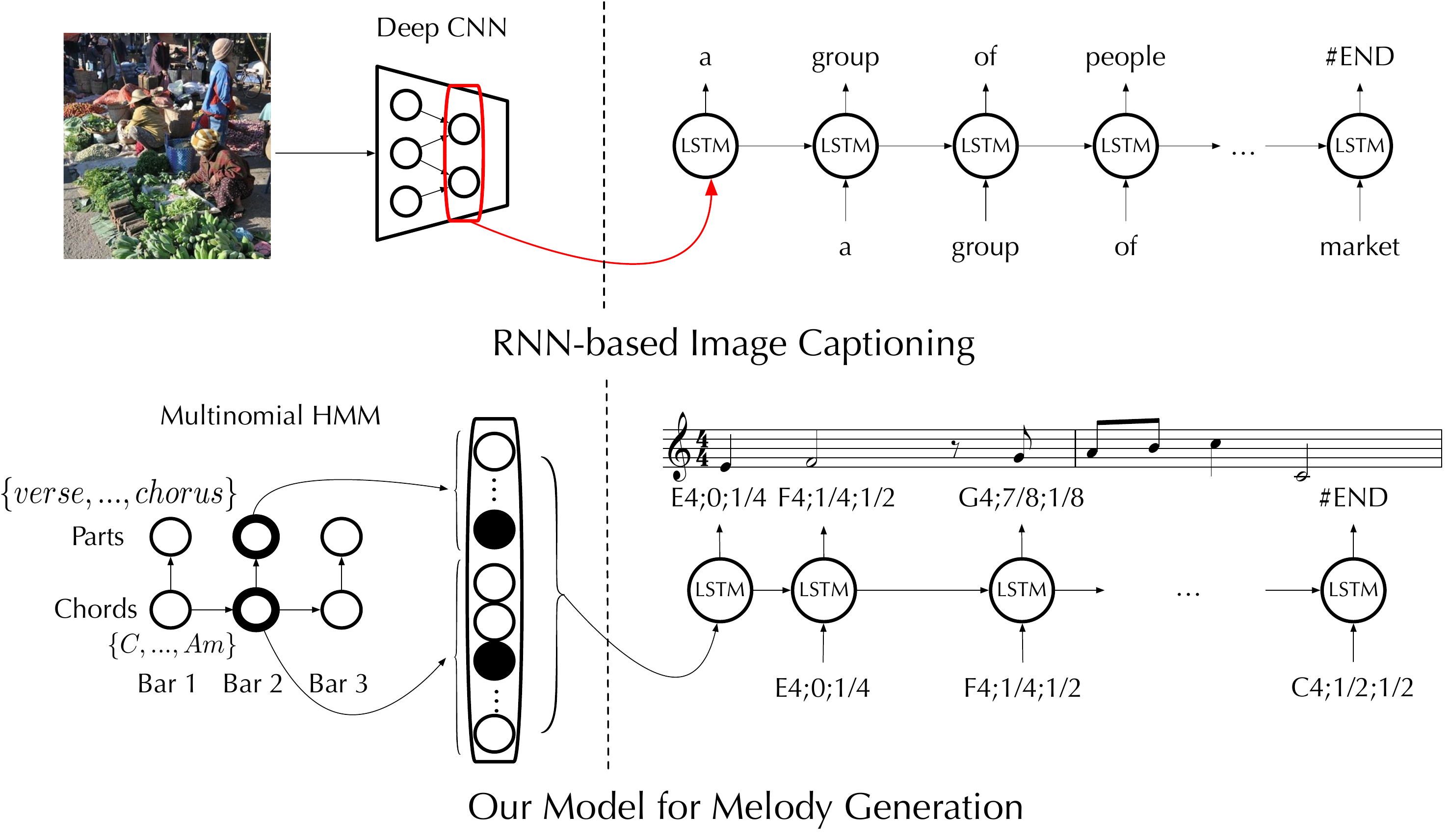}
	\caption{Visual analogy between image captioning task and our model. By grouping a note and its properties as a `\textit{word}', we generate melody as a `\textit{sentence}'.}
	\label{analogy}
	\vspace{-7mm}
\end{figure}

\section{Generation Model}
\subsection{Melody Representation}
Our model for melody generation can be best illustrated by making an analogy to image captioning task. In image captioning, most popular model is to generate each word sequentially via recurrent networks such as long short-term memory (LSTM) (\cite{LSTM}), conditioned on the image representation. In our model, we treat each note and its properties as a unique `word,' so that melody becomes the `sentence' to be generated. In other words, a pitch $p_i$ with duration $l_i$ located at $t_i$ within the current chord sequence will be represented as a single word $w_i=(p_i,t_i,l_i)$. Accordingly, a melody will be a sequence of \textit{words}, $s_j=(w_0,...,w_{m_i})\in S$. While we also use LSTM for word generation part, we condition it on music-relevant feature $x_i\in X$, instead of CNN image features; namely, chord-sequence $x_{chord_i}$ and part annotation $x_{part_i}$. Thus, we perform a maximum log likelihood estimation by finding the parameters set $\theta^*$ such that
\begin{equation}
\begin{aligned}
\theta^*=\arg\underset{\theta}{\max}\underset{(X,S)}{\sum}\log p(s_i|x_i;\theta)=\arg\underset{\theta}{\max}\sum_{1}^{N}\log p(w_0,...,w_{m_i}|x_{chord_i},x_{part_i};\theta)
\end{aligned}
\end{equation}
where \textit{N} is the number of training samples. Figure~\ref{analogy} makes a visual analogy between image captioning task and our model.

\begin{figure}[t] 
	\centering
	\subfloat[Independent representation of each property at identical time intervals (previous)]{%
		\includegraphics[width=0.48\textwidth]{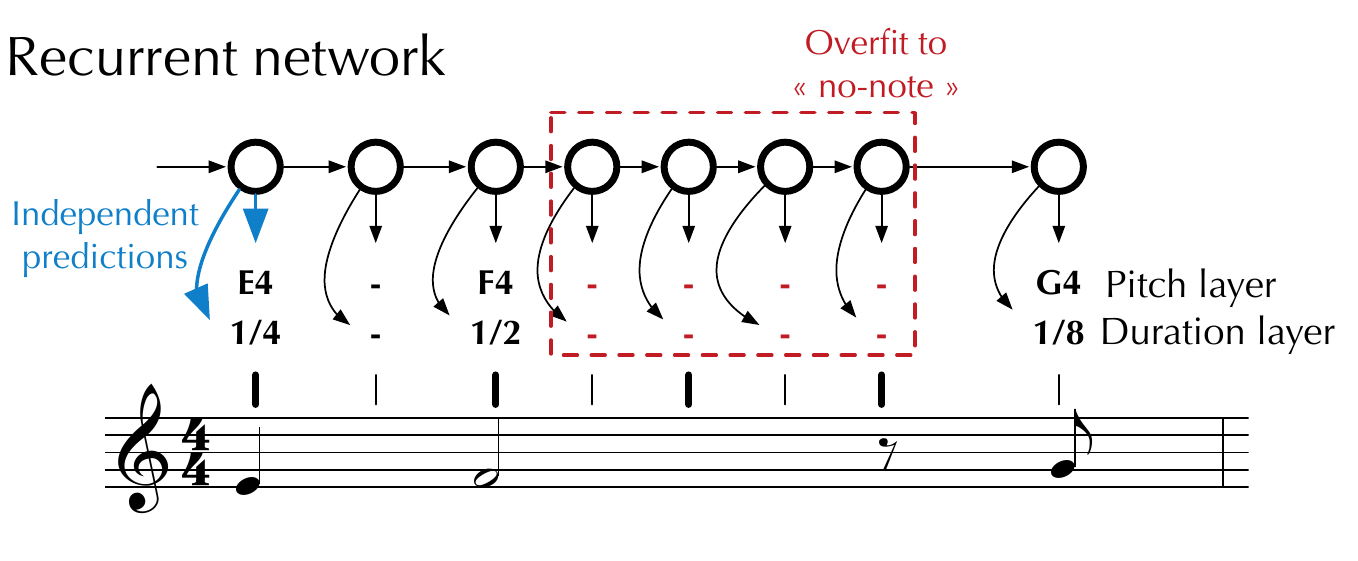}%
		\label{fig:a}%
	}%
	\hfill%
	\subfloat[Timestep-independent word representation of multiple properties (proposed)]{%
		\includegraphics[width=0.48\textwidth]{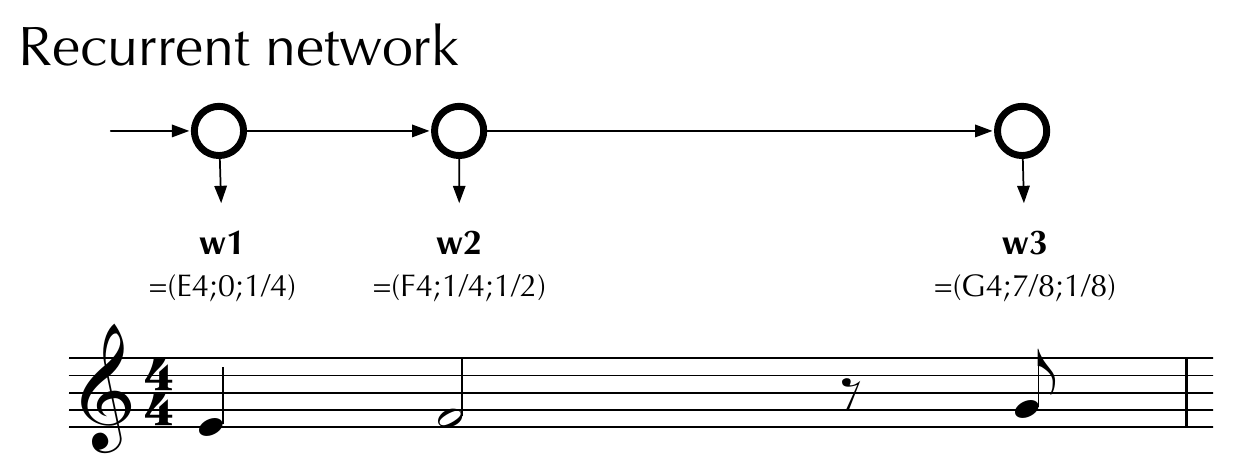}%
		\label{fig:b}%
	}%
	\caption{Comparison of our model for musical representation with previous model. Most of the previous models used a frame-level time granularity, which can easily lead to a model over-fitting on the repetition of the previous time step. Our proposed model of word representation alleviates this problem by encoding the time information (duration and position).}
	\vspace{-4ex}
	\label{comparison}
\end{figure}

Our model of melody representation makes a strong contrast with widely used approach of implementing separate layers for each property as described in Section 2. Previous approach essentially treats every 1/16 segment equally. Because this approach encounters a substantial number of segments that are repeated over several time steps, it is very likely that a statistically trained model will simply learn to repeat the previous segment, particularly for segments with no notes. It also complicates the learning by having to take the correlations among the different properties into consideration. On the other hand, our model does not have to consider intervals that do not contain notes, since our word representation already contains positional information. This puts us at advantage particularly when simultaneous notes are involved; even though notes are generated sequentially, they can be positioned at the same position, forming chords, which is difficult to implement with previous models based on time step. It also suffices to implement only one layer of generation, since the representation contains both pitch and length information. Moreover, considering pitch, position, and its length simultaneously is more concurrent with how humans would write melodies (\cite{Levitin}). Visual description of our model and previous model is shown in Figure~\ref{comparison}. 

Melody generation through outputting a sequence of `\textit{words}' is performed by LSTM with musical input features that will be described in Section 3.2. Following \cite{Karpathy}, word vectors were randomly initialized. We used the conventionally used gate functions for LSTM as following:
\begin{equation}
\begin{aligned}
i_t = \sigma(W_{i_x}x_t+W_{i_h}h_{t-1}+b_i)\\
f_t = \sigma(W_{f_x}x_t+W_{f_h}h_{t-1}+b_f)\\
o_t = \sigma(W_{o_x}x_t+W_{o_h}h_{t-1}+b_o)\\
g_t = tanh(W_{g_x}x_t+W_{g_h}h_{t-1}+b_g)
\end{aligned}
\end{equation}
where $\sigma$ indicates sigmoid function for non-linearity activation, $h_{t_1}$ is the memory output from the previous timestep that is fed back to LSTM, $b_i$ is bias, and $i_t$,$f_t$,$o_t$ correspond to input, foget, output gates respectively.


\subsection{Chord Sequence \& Part Generation}
Since our melody generation model is conditioned on musical input features, namely chord sequence and part information, we now examine how to automate the input generation part. We employ two-fold multinomial Hidden Markov Model (HMM), in which each chord and each part is a state whose state transition probabilities are computed according to our dataset. It works in a two-fold way, in which chord states are treated as latent variables whose transitions are dependent on the part states as observed variables. Thus,
\begin{equation}
p(x_1,...,x_m,z_1,...,z_N)=p(z_1)\prod_{n=2}^{N}p(z_N|z_{N-1})\prod_{n=1}^{N}p(x_n|z_N)
\end{equation}
where $x_i$ are part states and $z_i$ are chord states. Viterbi algorithm was used for decoding.

\subsection{Regularization}
Training with a large amount of data can lead the learning process to encounter a wide range of pitches, particularly when scale shifts are involved in the training data as in \cite{SongFromPi} or in our dataset. Such problem can lead to generation of unnatural melody whose pitch range deviates from what would be expected from a single song. We enforce regularization on the pitch range, so that the generated melody stays in a pitch range that humans would find easy to follow. We assign regularization cost to the learning process, so that a penalty is given in proportion to the absolute distance between the generated note and the nearest note in the pre-determined pitch range.
Algorithm 1 describes the procedure of our regularization on pitch range, whose outcome will be back-propagated to get gradients. We set minimum and maximum pitch as 60 (C4) and 72 (C5) respectively, but it can be easily adjusted depending on the desirable type of song or gender of target singer. We set regularization coefficient as 0.0001.

\makeatletter
\def\BState{\State\hskip-\ALG@thistlm}
\makeatother

\begin{algorithm}[t]
	\caption{Regularization for pitch range}\label{euclid}
	\begin{algorithmic}[1]
		\State \textbf{Inputs}: $W=$ initially empty weight matrix, $P=$ softmax predictions, $S$ = generated melody with pitches $(p_0,...,p_n)$, pre-set minimum and maximum pitches $p_{min}$ and $p_{max}$, coefficient $\mu$
		\State \textbf{for} $p_i$ \textbf{in} $S$
		\State \quad \textbf{if} $p_i > p_{min}$
		\State \qquad append $\max(p_i-p_{max},0)$ to $W$
		\State \quad \textbf{else} 
		\State \qquad append $p_{min}-p_i$ to $W$
		\State Sum up the products of $P$ and $W$ to get cost $C=\sum W_jP_j$
		\State Compute derivative $\frac{dE}{dp_i}=P_i(W_i-C)$
		\State Update softmax cost by adding $\frac{dE}{dp_i}\mu$
	\end{algorithmic}
\end{algorithm}
\vspace{-2ex}

\begin{table}[t]
	\small
	\begin{center}
		\caption{List of chord sequences over 2 continuous bars used in our dataset. Scale for all sequences has been adjusted to C Major.}
		\begin{tabular}{c}
			\hline \bf Chord Sequences  \\ \hline
			(C-Em), (A\#-F), (Dm-Em), (Dm-G), (Dm-C), (Am-Em), (F-C), (F-G), (Dm-F), (C-C), (C-E), (Am-G),\\
(F-F), (G-G), (Am-Am), (Dm-Dm), (C-A\#), (Em-F), (C-G), (G\#-A\#), (F-Am), ( G\#-Fm), (Am-Gm), (F-E),\\
(Dm-Am), (Em-Em), (G\#-G\#), (Em-Am), (C-Am), (F-Dm), (G\#-G), (F-A\#), (Am-G\#), (C-D), (G-Am),\\
(Am-C), (Am-A\#), (A\#-G), (Am-F), (A\#-Am), (E-Am), (Dm-E), (A-G), (Am-Dm), (Em-Dm), (C-F\#m),\\
(Am-D), (G\#-Em), (C-Dm), (C-F), (G-C), (A\#-A\#), (Am-Caug), (Fm-G), (A-A), (F-Em) \\ \hline

		\end{tabular}
		\label{cap_vs_narr_table}
	\end{center}
	\vspace{-5ex}
\end{table}

\section{Experiment}

\subsection{Setting}
\label{headings}
We collected 46 songs in MIDI format, most of which are unpublished materials from semi-professional musicians. Unofficial MIDI files for published songs were obtained on the internet, and we were granted the permission to use the songs for training from the organization owning the copyrights of the corresponding songs. It is very common in computer vision field to restrict a task to a certain domain so that the learning becomes more feasible. We also restricted our domain to pop music of major scale to make the learning more efficient.

Some of the previous works (\cite{DeepBach}) have employed data augmentation via scale shift. Instead, we adjusted each song's scale to C Major, thus eliminating the risk of mismatch between scale and generated melody. This adjustment has a side effect of widening the pitch range of melody beyond singable one, but this effect can be lessened by the regularization scheme over pitch range as described in Section 3.

We manually annotated chord and part for each bar in the songs collected. We restricted our chord annotation to only major and minor chords with one exception of C augmented \footnote{Note that, since all the songs have been adjusted to C major scale, we are using the tabular notation with root notes for convenience, instead of the conventional Roman numerals that are scale-invariant.}. Note, however, that this does not prevent the system from generating songs of more complex chords. For example, melodies in training data that are conditioned on C Major still contain notes other than the members of the conditioning triad, namely C, E, and G. Thus, our system may generate a non-member note, for example, B, as part of the generated melody when conditioned on C Major, thus indirectly forming C Major 7th chord. Part annotation consisted of 4 possible choices that are common in pop music structure; verse, pre-chorus, chorus, and bridge.

We experimented with n=1,2,4 continuous bars of chord sequences. Conditioning on only one bar generated melody that hardly displays any sense of continuity or theme. On the other hand, using chord progression over 4 bars led to data sparsity problem, which leads to generated songs simply copying the training data. Chord sequences over 2 bars thus became our natural choice, as it was best balanced in terms of both thematic continuity and data density. Check our demo for example songs conditioned on n=1,2,4 continuous bars. We annotated \textit{non-overlapping} chord sequences only; for example, given a sequence C - Am - F - G, we sample C - Am and F - G, but not the intermediate Am - F. This was our design choice to better retain the thematic continuity. As for the length of notes, we discretized by 16 if the length was less than 1/2, and by 8 otherwise.

Table~\ref{datastat} shows some of the statistics from our dataset. Throughout our dataset construction, pretty-midi \footnote{https://github.com/craffel/pretty-midi} framework was used to read, edit, and write MIDI files. Our dataset is publicly available with permissions from the owners of the copyright.

We ended up having 2082 unique `words' in our vocabulary. Learning rate was set to 0.001. Total number of learnable parameters was about 1.6M, and we applied dropout (\cite{Srivastava}) with 50\% probability after encoding to LSTM.

\begin{table}[t]
	\small
	\begin{center}
		\caption{Statistics of our dataset.}
		\label{datastat}
		\begin{tabular}{c|c|c|c|c|c}
			\hline   \bf \# songs &  \bf \# samples & \bf avg \# notes & \bf max \# notes & \bf min \# notes & \bf std dev \\ \hline
			46 & 1912 & 9.33 & 24  &1 & 3.60 \\ \hline
			\hline   \bf min pitch &  \bf max pitch & \bf median pitch & \bf min length & \bf max length & \bf median length \\ \hline
			53 & 86 & 69 & 1/16  &1 & 1/8 \\ \hline
			
		\end{tabular}
		\label{cap_vs_narr_table}
	\end{center}
	\vspace{-3ex}
\end{table}

\begin{figure}
	\centering
	{\includegraphics[width=3cm]{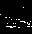}} 
	{\includegraphics[width=3cm]{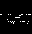}}
	{\includegraphics[width=3cm]{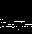}}
	{\includegraphics[width=3cm]{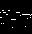}}
	\caption{Visualization of songs generated with GAN.} 
	\label{GAN_vis} 
	\vspace{-4ex}
\end{figure} 

\subsection{Evaluation}
We make comparison to some of the recent works that employed deep learning to generate music in MIDI format. We performed two kinds of human evaluation tasks on Amazon Mechanical Turk, making comparison between outcomes from our model and two baseline models; \cite{SongFromPi} and \cite{Magenta1}. We deliberately excluded \cite{DeepBach} as it belongs to a different domain of classical music. In task 1, we first asked the participants how much expertise they have in music. We then played one song from our model and another song from one of the baseline models. After listening to both songs, participants were asked to answer which song has melody that sounds more like human-written one, which song is more well-structured, and which one they like better. In task 2, we performed a type of Turing test (\cite{Turing}) in which the participants were asked to determine whether the song was written by human or AI.

Table~\ref{task1} shows the results from task 1 for each question and each expertise level of the participants. 973 workers participated. Against \cite{SongFromPi}, our model was preferred in all aspects, suggesting our model's superiority over their multi-layer generation. Against \cite{Magenta1}, our model was preferred in all aspects except in structure. Lower score in structure is most likely due to their musical formality enabled by pre-defined set of theoretical rules. Yet, our model, without any pre-defined rule, was considered to have more natural melodies and was more frequently preferred. Interestingly, even when participants determined that one song has more human-like melody with clearer structure, they frequently answered that they preferred the other song, implying that human-ness may not always correlate to musical taste. $\chi^2$ statistic is 75.69 against \cite{SongFromPi} and 31.17 against \cite{Magenta1}, with \textit{p}-value less than 1e-5 in both cases. Against either baseline model, people with intermediate or high expertise in music tended to prefer our model than those with low expertise. Table~\ref{task2} shows the results from task 2. 986 workers participated. Understandably, songs actually written by humans had the largest proportion of being judged as humans. Our model had the best deception rate among the artificial generation models. Consistency of the results with task 1 implies that generating natural melody while preserving structure is a key for human-like music generation.

\begin{table}[t]
	\small
	\begin{center}
		\caption{Results from evaluation task 1. Numbers indicate the proportion in which our model was preferred over the baseline model.}
		\label{task1}
		\begin{tabular}{c|c|c|c|c|c}
			\hline \bf vs. Model & \bf Expertise & \bf Melody  &  \bf Structure & \bf Preference & \bf Overall \\ \hline
			\multirow{4}{*}{vs. \cite{SongFromPi}} &low & .598  & .545 & .542 & .576 \\ \cline{2-6}
			&middle& .719  & .692 & .619 & .684 \\ \cline{2-6}
			&high& .687 & .712 & .712 & .704   \\ \cline{2-6}
			&all& .691 & .684 & .654 & .678 \\ \hline
			\multirow{4}{*}{vs. \cite{Magenta1}}&low & .583  & .458  &  .583  &  .542  \\ \cline{2-6}
			&middle& .570 &.427  &  .567 &  .521  \\ \cline{2-6}
			&high& .598 &  .511  &  .565 &  .558\\ \cline{2-6}
			&all& .577 & .447 & .567  & .530 \\ \hline
		\end{tabular}
	\end{center}
	\vspace{-2ex}
\end{table}

\begin{table}
	\small
	\begin{center}
		\caption{Results from evaluation task 2. Deception rate indicates the proportion in which the song was believed to be made by human.}
		\label{task2}
		\begin{tabular}{c|c|c|c|c}
			\hline \bf Model & \bf Ours & \bf  \cite{SongFromPi} &\bf  \cite{Magenta1} & \bf Human \\ \hline
			\bf Deception rate & .680 & .599 & .620 & .777 \\ \hline
		\end{tabular}
		\label{cap_vs_narr_table}
	\end{center}
	\vspace{-5ex}
\end{table}

\subsection{Additional Experiments}
Generative Adversarial Networks (GANs) (\cite{GAN}) have proven to be a powerful technique to generate visual contents, to the extent where the generated results are frequently indistinguishable from human-made contents or actual pictures (\cite{Radford}; \cite{reed2016}). Since the musical score can be regarded as a one-dimensional image with the time direction as the x axis and the pitch as the channel, we hypothesized that GAN may be able to generate music as \textit{image}.

GANs consist of a generator $G$ and a discriminator $D$. The generator $G$ receives random noise $z$ and condition $c$  as inputs, and outputs contents $G(z, c)$. The discriminator $D$ distinguishes between real data $x$ in the dataset and the outputs of the generator $G(z, c)$. The discriminator $D$ also receives the condition $c$. $D$ is trained to minimize $-\log(D(x, c)) - \log(1 - D(G(x, c), c))$ while $G$ is trained to minimize $-\log(D(G(x, c), c))$. We used the two-hot feature vector described in Section 3 as condition $c$. We used down-sampling \& up-sampling architecture, Adam optimizer (\cite{Adam}), and batch normalization (\cite{BatchNorm}) as suggested in \cite{Radford}. 

Listening to the generated results, it does have its moments, but is frequently out of tune and the melody patterns sound restricted. GAN does have advantage particularly with chords, as it can `visually' capture the harmonies, as opposed to sequential generation in our proposed model, or stacking different layers of single notes as in \cite{DeepBach}. Also, melody generation with GAN can potentially avoid the problem of overfitting due to elongated training. On the other hand, the same melody frequently appears for the same input. This is likely due to the problem known as GAN's mode collapse, in which the noise input is mostly ignored. In addition, it is difficult to know whether a line corresponds to a single note or consecutive notes of smaller lengths. Many of the problems seem to fundamentally stem from the difference in modalities; image and music. See Figure~\ref{GAN_vis} for visualization of the songs generated with GAN.

We also examined generating other instrument tracks on top of the melody track using the proposed model. We extracted bass tracks, piano tracks, and string tracks from the dataset, and performed the same training procedure as described in Section 3. Generated instruments sound fairly in tune individually, confirming that our proposed model is applicable to other instruments as well. Moreover, we were able to generate instrument tracks with simultaneous notes (chords), which is difficult to implement with previous generation model based on time step. However, combining the generated instrument tracks to make a 4-track song resulted in dissonant and unorganized songs. This implies that generating a multi-track song requires a more advanced model for learning that reflects the interrelations among the instruments, which will be our immediate future work. Check our demo for songs generated with GAN and multi-track song generated with our model.

\subsection{Discussion}
Although our model was inspired by the model used in image captioning task, its task objective has a fundamental difference from that of image captioning. In image captioning task, more resemblance to human-written descriptions reflects better performance. In fact, matching human-written descriptions is usually the evaluation scheme for the task. However, in melody generation, resembling human-written melody beyond certain extent becomes \textit{plagiarism}. Thus, while we need sufficient amount of training to learn the patterns, we also want to avoid overfitting to training data at the same time. This poses questions about how long to train, or essentially how to design the loss function. We examined generations with parameters learned at different epochs. Generated songs started to stay in tune roughly after 5 epochs. However, after 20 epochs and on, we could frequently observe the same melodies as in the training data, implying overfitting (check our demo). So there seems to exist a `\textit{safe zone}' in which it learns enough from the data but not exceedingly to copy it. Previous approaches like \cite{Magenta1} have dealt with this dilemma by rewarding for following the pre-determined rules, but encouraging off-policy at the same time. Since we aim for learning without pre-determined rules, alternative would be to design a loss function where matching the melody in training data over \textit{n} consecutive notes of threshold is given penalty. Designing a more appropriate loss function remains as our future work. On the other hand, generating songs with parameters obtained at different stages within the `safe zone' of training leads to diversity of melodies, even when the input vectors are identical. This property nicely complements our relatively low-dimensional input representation.

\section{Conclusion \& Future Works}
In this paper, we proposed a novel model to generate melody for pop music. We generate melody with word representation of notes and their properties, instead of training multiple layers for each property, thereby reducing the complexity of learning. We also proposed a regularization model to control the outcome. Finally, we implemented part-dependent melody generation which helps the generated song preserve the overall structure, along with a publicly available dataset. Experimental results demonstrate that our model can generate songs whose melody sounds more like human-written ones, and is more well-structured than previous models. Moreover, people found it more difficult to distinguish the songs from our model from human-written songs than songs from previous models. On the other hand, examining other styles such as music of minor scale, or incorporating further properties of notes, such as intensity or vibrato, has not been examined yet, and remains as future work. As discussed in Section 4, learning to model the correlations among different instruments also remains to be done, and designing an appropriate loss function for the task is one of the most critical tasks to be done. We plan to constantly update our dataset and repository, addressing the future works.
\label{others}

\subsubsection*{Acknowledgments}
We thank the Japanese Society for Rights of Authors, Composers, and Publishers (JASRAC) for granting us permissions to use the songs for training.

\bibliography{iclr2018_conference}

\begin{thebibliography}{21}
\providecommand{\natexlab}[1]{#1}
\providecommand{\url}[1]{\texttt{#1}}
\expandafter\ifx\csname urlstyle\endcsname\relax
  \providecommand{\doi}[1]{doi: #1}\else
  \providecommand{\doi}{doi: \begingroup \urlstyle{rm}\Url}\fi

\bibitem[Chu et~al.(2017)Chu, Urtasun, and Fidler]{SongFromPi}
Hang Chu, Raquel Urtasun, and Sanja Fidler.
\newblock Song from {PI:} {A} musically plausible network for pop music
  generation.
\newblock In \emph{ICLR Workshop}, 2017.

\bibitem[Gal \& Ghahramani(2016)Gal and Ghahramani]{gal}
Yarin Gal and Zoubin Ghahramani.
\newblock Bayesian convolutional neural networks with bernoulli approximate
  variational inference.
\newblock In \emph{ICLR Workshop}, 2016.

\bibitem[Goodfellow et~al.(2014)Goodfellow, Pouget-Abadie, Mirza, Xu,
  Warde-Farley, Ozair, Courville, and Bengio]{GAN}
Ian Goodfellow, Jean Pouget-Abadie, Mehdi Mirza, Bing Xu, David Warde-Farley,
  Sherjil Ozair, Aaron Courville, and Yoshua Bengio.
\newblock Generative adversarial nets.
\newblock In \emph{NIPS}, 2014.

\bibitem[Hadjeres \& Pachet(2017)Hadjeres and Pachet]{DeepBach}
Ga{\"{e}}tan Hadjeres and Fran{\c{c}}ois Pachet.
\newblock Deepbach: a steerable model for bach chorales generation.
\newblock In \emph{ICML}, 2017.

\bibitem[Hochreiter \& Schmidhuber(1997)Hochreiter and Schmidhuber]{LSTM}
Sepp Hochreiter and J\"{u}rgen Schmidhuber.
\newblock Long short-term memory.
\newblock \emph{Neural Computation}, 9\penalty0 (8):\penalty0 1735--1780, 1997.

\bibitem[Huang et~al.(2017)Huang, Cooijmans, Roberts, Courville, and
  Eck]{Magenta2}
Cheng-Zhi~Anna Huang, Tim Cooijmans, Adam Roberts, Aaron Courville, and Douglas
  Eck.
\newblock Counterpoint by convolution.
\newblock In \emph{ISMIR}, 2017.

\bibitem[Ioffe \& Szegedy(2015)Ioffe and Szegedy]{BatchNorm}
Sergey Ioffe and Christian Szegedy.
\newblock Batch normalization: Accelerating deep network training by reducing
  internal covariate shift.
\newblock In \emph{ICML}, 2015.

\bibitem[Jacob(1996)]{rule1}
Bruce~L. Jacob.
\newblock Algorithmic composition as a model of creativity.
\newblock \emph{Organised Sound}, 1\penalty0 (3), 1996.

\bibitem[Jaques et~al.(2017)Jaques, Gu, Turner, and Eck]{Magenta1}
Natasha Jaques, Shixiang Gu, Richard~E. Turner, and Douglas Eck.
\newblock Tuning recurrent neural networks with reinforcement learning.
\newblock In \emph{ICLR Workshop}, 2017.

\bibitem[Karpathy \& Li(2015)Karpathy and Li]{Karpathy}
Andrej Karpathy and Fei-Fei Li.
\newblock {Deep Visual-Semantic Alignments for Generating Image Descriptions}.
\newblock In \emph{CVPR}, 2015.

\bibitem[Kingma \& Ba(2015)Kingma and Ba]{Adam}
Diederik~P. Kingma and Jimmy Ba.
\newblock Adam: {A} method for stochastic optimization.
\newblock In \emph{ICLR}, 2015.

\bibitem[Levitin(2006)]{Levitin}
D.J. Levitin.
\newblock \emph{This is Your Brain on Music: The Science of a Human Obsession}.
\newblock Dutton, 2006.

\bibitem[Papadopoulos \& Wiggins(1999)Papadopoulos and Wiggins]{rule2}
George Papadopoulos and Geraint Wiggins.
\newblock Ai methods for algorithmic composition: A survey, a critical view and
  future prospects.
\newblock In \emph{AISB Symphosium on Musical Creativy}, 1999.

\bibitem[Potash et~al.(2015)Potash, Romanov, and Rumshisky]{ghost}
Peter Potash, Alexey Romanov, and Anna Rumshisky.
\newblock Ghostwriter: Using an {LSTM} for automatic rap lyric generation.
\newblock In \emph{EMNLP}, 2015.

\bibitem[Radford et~al.(2016)Radford, Metz, and Chintala]{Radford}
Alec Radford, Luke Metz, and Soumith Chintala.
\newblock Unsupervised representation learning with deep convolutional
  generative adversarial networks.
\newblock In \emph{ICLR}, 2016.

\bibitem[Reed et~al.(2016)Reed, Akata, Mohan, Tenka, Schiele, and
  Lee]{reed2016}
Scott Reed, Zeynep Akata, Santosh Mohan, Samuel Tenka, Bernt Schiele, and
  Honglak Lee.
\newblock Learning what and where to draw.
\newblock In \emph{NIPS}, 2016.

\bibitem[Srivastava et~al.(2014)Srivastava, Hinton, Krizhevsky, Sutskever, and
  Salakhutdinov]{Srivastava}
Nitish Srivastava, Geoffrey Hinton, Alex Krizhevsky, Ilya Sutskever, and Ruslan
  Salakhutdinov.
\newblock Dropout: A simple way to prevent neural networks from overfitting.
\newblock \emph{J. Mach. Learn. Res.}, 15\penalty0 (1), 2014.

\bibitem[Turing(1950)]{Turing}
A.~M. Turing.
\newblock Computing machinery and intelligence.
\newblock \emph{Mind}, 59\penalty0 (236), 1950.

\bibitem[van~den Oord et~al.(2016)van~den Oord, Dieleman, Zen, Simonyan,
  Vinyals, Graves, Kalchbrenner, Senior, and Kavukcuoglu]{WaveNet}
A{\"{a}}ron van~den Oord, Sander Dieleman, Heiga Zen, Karen Simonyan, Oriol
  Vinyals, Alex Graves, Nal Kalchbrenner, Andrew~W. Senior, and Koray
  Kavukcuoglu.
\newblock Wavenet: {A} generative model for raw audio.
\newblock \emph{CoRR}, abs/1609.03499, 2016.

\bibitem[Vinyals et~al.(2015)Vinyals, Toshev, Bengio, and Erhan]{ShowAndTell}
Oriol Vinyals, Alexander Toshev, Samy Bengio, and Dumitru Erhan.
\newblock {Show and Tell: A Neural Image Caption Generator}.
\newblock In \emph{CVPR}, 2015.

\bibitem[Xu et~al.(2015)Xu, Ba, Kiros, Cho, Courville, Salakhudinov, Zemel, and
  Bengio]{ShowAttendTell}
Kelvin Xu, Jimmy Ba, Ryan Kiros, Kyunghyun Cho, Aaron Courville, Ruslan
  Salakhudinov, Richard Zemel, and Yoshua Bengio.
\newblock {Show, Attend and Tell: Neural Image Caption Generation with Visual
  Attention}.
\newblock In \emph{ICML}, 2015.

\end{thebibliography}
\bibliographystyle{iclr2018_conference}

\end{document}